\documentclass{article}
\usepackage{inputenc,graphicx,amsmath,caption,float,amssymb,csquotes}

\MakeOuterQuote{"}

\title{Take the $A$-Metric: Interpretations of Some Known Solutions of Einstein's Vacuum Field Equations}
\author{Charles W. Robson* and Marco Ornigotti}
\date{}

\begin{document}

\maketitle

\vspace{-10mm}
\begin{center}
Laboratory of Photonics, Physics Unit, Tampere University, FI-33720 Finland
*email: charles.robson@tuni.fi
\end{center}
\vspace{5mm}

\noindent In this work, we present a new interpretation of the only static vacuum solution of Einstein's field equations with planar symmetry, the Taub solution. This solution is a member of the $AIII$ class of metrics, along with the type D Kasner solution. Various interpretations of these solutions have been put forward previously in the literature, however, some of these interpretations have suspect features and are not generally considered physical. Using a simple mathematical analysis, we show that a novel interpretation of the Taub solution is possible and that it naturally emerges from the radial, near-singularity limit of negative-mass Schwarzschild spacetime. A new, more transparent derivation is also given showing that the type D Kasner metric can be interpreted as a region of spacetime deep within a positive-mass Schwarzschild black hole. The dual nature of this class of $A$-metrics is thereby demonstrated. \\


\section{Introduction}

Over the previous century, a vast number of solutions of Einstein's field equations have been discovered \cite{Stephani,Ehlers,MTW,Wald,Carroll,Petrov_book,Belinski}. During this same period, however, a much smaller number of agreed-upon physical interpretations of these solutions have been established \cite{Griffiths,Bonnor1,Bonnor2,Bonnor2_2,Zhang}. Some spacetime solutions have clear interpretations: there is a consensus, for example, that certain metrics describe black holes, gravitational waves, and cosmologies, but other solutions have much more mysterious origins -- some can be interpreted as describing \emph{local} regions of realistic spacetimes, some indeed have multiple feasible readings, and others currently have none at all \cite{Griffiths}. Dealing with the protean nature of solutions, a by-product of the general covariance of the field equations, can require subtle methods such as obscure coordinate transformations \cite{Harvey}.

Hidden within the structure of general relativity, there are doubtless undiscovered solutions describing novel spacetime configurations with relevance and importance today. It is nevertheless arguable that the immediate task is to interpret those solutions that have already been found through many years of great effort \cite{Stephani,Griffiths,Ehlers}. It is a pressing task then, as it has been since the early days of general relativity, to find reasonable and useful interpretations of all its solutions and to reassess those with doubtful features.

In this paper, we make a contribution to this task by presenting a new interpretation of a class of solutions known as the $AIII$-metrics \cite{Griffiths,Ehlers}. Previous explanations of the physics described by these solutions contain questionable properties, including fine-tuned parameters \cite{Griffiths}. We argue that our new interpretation is by far the most natural yet suggested, relating as it does to a local region of the Schwarzschild solution with finite mass parameter. Our result is found using only simple mathematical analysis.

The $AIII$-metrics are a subclass of the more general $A$-metrics family, the latter classified by Ehlers and Kundt in 1962 \cite{Ehlers}. Each subclass of the $A$-metrics family is characterised by a 2-space curvature parameter. Depending on the choice of this parameter, one finds the well-known Schwarzschild solution ($AI$), a solution with a negative-Gaussian-curvature hypersurface ($AII$), capable of describing the gravitational field produced by a tachyon \cite{Peres}, or the $AIII$-metrics;
in this paper, we focus purely on the final subclass, as the different interpretations and geometrical features present (including a curvature singularity) in these vacuum solutions make them an interesting topic of study.

As has been previously stated \cite{Griffiths}, "most solutions of Einstein's equations may have no satisfactory interpretation as global models of realistic physical situations at all. On the other hand, locally, many may reasonably represent particular regions of realistic space-times". Consistent with this point of view, it is local spacetime regions that are the focus of this paper.

In Sec. \ref{sec:results}, we present our results reinterpreting the $AIII$-metrics as local regions of Schwarzschild spacetime. Sec. \ref{sec:discussion} contains a critical survey of previous readings of the $AIII$-metrics, putting forward the case that ours is the most natural, avoiding as it does characteristics that render other interpretations unphysical or obscure. Sec. \ref{sec:conclusions} contains a summary of the paper. Several appendices contain useful coordinate transformations, a discussion of how the Kasner metric relates to the physics inside a black hole, and a description of the geometry connected with our results.

\section{Results} \label{sec:results}

The family of vacuum spacetimes known as the \emph{$A$-metrics} will be the focus of this work and is given by \cite{Ehlers,Griffiths}
\begin{equation} \label{eq:A_metrics}
ds^2 = -\left( \epsilon - \frac{2M}{r} \right) dt^2 + \left( \epsilon - \frac{2M}{r} \right)^{-1} dr^2 + r^2 \frac{2d\zeta d\bar{\zeta}}{\left( 1 + \epsilon \zeta \bar{\zeta}/2 \right)^2} .
\end{equation}
These are a rich and intriguing generalisation of the Schwarzschild solution, describing different spacetimes depending on the value of parameter $\epsilon$ (which can be $+1$, $-1$ or $0$); fixing the value of $\epsilon$ determines the Gaussian curvature of the hypersurface on which $t$ and $r$ are constant. Setting $\epsilon=1$ produces the $AI$-metric, i.e. the Schwarzschild solution. $M$ is an arbitrary, continuous parameter and $\bar{\zeta}$ is the complex conjugate of $\zeta$. Relevant coordinate ranges will be discussed below.

In this paper, we restrict our attention to the \emph{AIII-metrics}, which emerge setting $\epsilon=0$ in Eq. (\ref{eq:A_metrics}). Performing the coordinate transformation $\zeta = \rho \mathrm{exp}(i\phi)/\sqrt{2}$ with $\phi \in [ 0 , 2\pi )$ then gives
\begin{equation} \label{eq:AIII_metrics}
ds^2 = \frac{2M}{r} dt^2 - \frac{r}{2M} dr^2 + r^2 \left( d\rho^2 + \rho^2 d\phi^2 \right) .
\end{equation}
Due to the curvature singularity at $r=0$, it is appropriate to constrain the value of the coordinate as $r>0$. The line element (\ref{eq:AIII_metrics}) has been interpreted in various different ways in the literature, as will be discussed in detail in Sec. \ref{sec:discussion}.

When $M$ is positive and finite, Eq. (\ref{eq:AIII_metrics}) describes the polar spacetime deep within a Schwarzschild black hole of finite positive mass. What has not been noted before, is that when $M$ is \emph{negative} and finite, metric (\ref{eq:AIII_metrics}) describes the small-$r$, polar region of negative-finite-mass Schwarzschild spacetime. By "polar spacetime" we mean that the spacetime is restricted to $\theta\ll1$, where $\theta$ is the polar angle of the Schwarzschild coordinate system (see below), and "deep within" a Schwarzschild black hole denotes $r\ll2M$, where $r$ is the radial Schwarzschild coordinate. Note that the negative-mass case has no event horizon and by "small-$r$" we signify $r\ll2|M|$, with $2|M|$ the characteristic length scale of the system. As there is no horizon for the negative-mass Schwarzschild spacetime, it contains a naked singularity; for a discussion of the physical feasibility of this configuration, see Ref. \cite{Gleiser} and references therein. For the rest of this paper, we refer to the zone $\theta\ll1$, $r\ll2|M|$ as the \emph{deep-radial region}.

New derivations supporting the above statements are now given. Starting from the Schwarzschild solution with $M>0$ \cite{Carroll},
\begin{equation} \label{eq:Schwarz}
ds^2 = -\left( 1 - \frac{2M}{r} \right) dt^2 + \left( 1 - \frac{2M}{r} \right)^{-1} dr^2 + r^2 \left( d\theta^2 + \mathrm{sin}^2 \theta d\phi^2 \right) ,
\end{equation}
and taking the limits $r\ll 2M$ and $\theta\ll 1$, one can see that the term $1-2M/r$ can be approximated by $-2M/r$ and that $\mathrm{sin}^2 \theta \approx \theta^2 + \mathcal{O}(\theta^4)$. Because $\theta\ll1$, we discard the $\mathcal{O}(\theta^4)$ term in the Taylor series of $\mathrm{sin}^2 \theta$, leading to the metric
\begin{equation} \label{eq:metric_pos_M}
ds^2 \approx \frac{2M}{r} dt^2 - \frac{r}{2M} dr^2 + r^2 \left( d\theta^2 + \theta^2 d\phi^2 \right) , \hspace{0.5cm} M>0.
\end{equation}

In general relativity, the coordinates have no intrinsic physical meaning and must be interpreted \cite{MTW}. We are therefore free to interpret $\theta$ in Eq. (\ref{eq:metric_pos_M}) as a dimensionless radial coordinate, which, together with azimuth $\phi$, defines a two-dimensional polar coordinate system. This coordinate reinterpretation after taking the limits $r\ll 2M$ and $\theta\ll 1$ above is natural, as the line element section $d\theta^2 + \theta^2 d\phi^2$ from (\ref{eq:metric_pos_M}) is intrinsically flat. A relabelling $\theta \rightarrow \rho$ in (\ref{eq:metric_pos_M}) yields the $AIII$-metric (\ref{eq:AIII_metrics}) with $M>0$. The $t$ coordinate spans the real line.

In case the reinterpretation of coordinates above seems obscure, all we have done is to approximate the surface of a two-sphere as \emph{locally} Euclidean (when $\theta \ll 1$ holds). We are free to do this as the two-sphere $S^2$ is a manifold and so, by definition, locally "looks like" $\mathbb{R}^2$ \cite{Carroll}. If one chose to, one could "sew together" these pieces which locally resemble $\mathbb{R}^2$ in order to reconstruct $S^2$ \cite{Wald}. Each of us unconsciously makes this approximation every day -- the Earth appears locally flat as one walks on its surface even though it is of course globally (roughly) spherical.

An important point concerning the $\theta \ll 1$ constraint is that it does not pick out any special pole in Schwarzschild spacetime, thereby breaking the solution's spherical symmetry \cite{Lambourne}. The choice of the origin of the $\theta$ coordinate is arbitrary (in other words, there is no "north pole" of a Schwarzschild black hole) and fixing $\theta \ll 1$ simply restricts the geometry to a thin cone of the spacetime, centred around any radial line. The additional constraint $r \ll 2M$ then "places" this cone behind the horizon. Note that the coordinate $r$ is \emph{timelike} in the deep-radial region for $M>0$ and this must be kept in mind when visualising the spacetime. See \textbf{Appendix A} for details.

The $AIII$-metric with $M>0$ can thus be interpreted as a local description of Schwarzschild spacetime in the limits $r\ll 2M$ and $\theta\ll 1$, using a simple approximation.

The preceding analysis held for $M>0$. In the case of the Schwarzschild solution with negative mass $M<0$, the metric can be written as
\begin{equation}
ds^2 = -\left( 1 + \frac{2|M|}{r} \right) dt^2 + \left( 1 + \frac{2|M|}{r} \right)^{-1} dr^2 + r^2 \left( d\theta^2 + \mathrm{sin}^2 \theta d\phi^2 \right) .
\end{equation}
Taking the limits $r\ll 2|M|$ and $\theta\ll 1$, the term $1+2|M|/r$ can be approximated by $2|M|/r$, giving
\begin{equation} \label{eq:metric_mod_M}
ds^2 \approx -\frac{2|M|}{r} dt^2 + \frac{r}{2|M|} dr^2 + r^2 \left( d\theta^2 + \theta^2 d\phi^2 \right) .
\end{equation}
Substituting $M=-|M|$ into metric (\ref{eq:metric_mod_M}) and relabelling $\theta \rightarrow \rho$ as before then yields
\begin{equation} \label{eq:metric_neg_M}
ds^2 \approx \frac{2M}{r} dt^2 - \frac{r}{2M} dr^2 + r^2 \left( d\rho^2 + \rho^2 d\phi^2 \right) , \hspace{0.5cm} M<0.
\end{equation}

Metrics (\ref{eq:metric_pos_M}) and (\ref{eq:metric_neg_M}) are therefore just the positive- and negative-"mass" branches\footnote{An equivalent mathematical viewpoint is that instead of looking at the $AIII$-metrics for $M>0$ and $M<0$ separately, both having coordinate range $r>0$, one can fix $M$ as positive and expand the radial domain to $r<0$ and $r>0$. This was also noted in Ref. \cite{Bedran}. Whether this has any physical significance is unknown to the authors.} of the $AIII$-metrics (\ref{eq:AIII_metrics}). The coordinate domain of $\phi$ is the same throughout: $\phi \in [ 0 , 2\pi )$. Of course, due to the domain constraints introduced, in Eq. (\ref{eq:AIII_metrics}) the upper limit of the $\rho$ coordinate must be restricted to a small, finite number $\rho_{0}$, i.e. $\rho \in [ 0 , \rho{_0} \ll 1 ]$, and the $r$ coordinate has to obey $0 < r \ll 2|M|$.

Another way to compare geometries is via the use of curvature invariants. The following scalar invariants, defined using the metric tensor, are equal for both the ($M>0$) $AIII$- and Schwarzschild metrics: the Ricci scalar ($R=0$) and Kretschmann scalar ($K=48M^2 / r^6$) \cite{Carroll,Ruffini}. Another invariant, the Karlhede scalar $K_2$ \cite{Karlhede,Ong_thesis}, differs globally over $r$: for Schwarzschild spacetime, $K_2 = 720M^2 (r-2M) / r^9$, whereas for the $AIII$ class of solutions, $K_2 = -1440M^3 / r^9$. It is clear however that in the limit $r \ll 2M$ the Karlhede scalars approximately match, as expected. The matching becomes closer as the singularity is approached. This argument can be easily adapted to the $M<0$ case, showing a matching there too.

\section{Discussion} \label{sec:discussion}

There exist in the literature several different interpretations of the $AIII$-metrics, with various properties. We introduce and discuss them in this section, comparing them critically with ours.

Applying a simple coordinate transformation to the $AIII$-metric (\ref{eq:AIII_metrics}) for the case $M>0$ demonstrates that it is the type D vacuum Kasner solution \cite{Griffiths} (for more details on this metric, and the vacuum Bianchi I class of solutions to which it belongs, see Refs. \cite{Kasner1921,Kasner1925,Harvey,Ellis}). This solution can be interpreted as a description of a certain anisotropic cosmology beginning with a (spacelike) big bang-type singularity \cite{Griffiths}. This "cosmological picture" (as we refer to it from here on) does not clash with the interpretation of the Kasner solution as a description of the deep-radial region of a Schwarzschild black hole\footnote{The link between the Kasner and Schwarzschild solutions has been noted before, see e.g. Refs. \cite{Frolov,Frolov2,Hiscock,Ashtekar,Matyjasek}.}. Both are valid, as we now show.

Performing the aforementioned coordinate transformation (see \textbf{Appendix B} for details) on the $M>0$ $AIII$-metric (\ref{eq:AIII_metrics}) yields the line element
\begin{equation} \label{eq:Kasner}
ds^2 = -d\tau^2 + \tau^{-2/3} dz^2 + \tau^{4/3}(dx^2 + dy^2) .
\end{equation}
The cosmological picture asserts that the type D vacuum Kasner solution describes a spacetime originating in a spacelike singularity at the origin $\tau=0$ [equivalent to $r=0$ in Eq. (\ref{eq:AIII_metrics})]. The coordinate origin $r=0$ is also a spacelike singularity in Schwarzschild spacetime, hence both interpretations account for this feature\footnote{The similar presence of essential singularities in the Kasner and Schwarzschild solutions was noted by Aichelberg in Ref. \cite{Aichelburg}.}. The fact that the coordinate $t$ is spacelike and $r$ is timelike for the $AIII$-metric (\ref{eq:AIII_metrics}) with $M>0$ is interpreted in the cosmological picture by treating $r$ as related to a global time coordinate [i.e. $\tau$ in Eq. (\ref{eq:Kasner})] characterising the evolution of the spacetime, and by taking $t$ as proportional to a spatial dimension [i.e. $z$ in Eq. (\ref{eq:Kasner})]. The deep-radial interpretation also explains these features, as the Schwarzschild coordinates $t$ and $r$ are spacelike and timelike, respectively, behind the horizon.

In the cosmological picture, the vacuum spacetime (after its "big bang" at $\tau=0$) expands uniformly in two directions whilst contracting in the remaining spatial dimension. As was shown in Sec. \ref{sec:results}, the deep-radial region of Schwarzschild spacetime can be approximated by a flat disk for $(t,r)=$constant. If one looks at the geometry of this disk, one can see how it matches the geometry in the cosmological picture, supplying a parallel viewpoint. Let us focus our attention on one piece of metric (\ref{eq:metric_pos_M}), specifically
\begin{equation} \label{eq:growing_disk_sector}
ds^2 \big|_{\left( \theta , \phi \right)} = r^2 \left( d\theta^2 + \theta^2 d\phi^2 \right) ,
\end{equation}
where the notation $ds^2 \big|_{\left( \theta , \phi \right)}$ denotes the $\left( \theta , \phi \right)$ sector of the line element. Our approximation demands $\theta\ll 1$, so we restrict the coordinate $\theta$ to the range $\theta \in [0,\theta_0]$, where $\theta_0\ll 1$. As mentioned earlier, the symbol $\theta$ is open to interpretation, as are all coordinates in general relativity, and the part of the metric (\ref{eq:growing_disk_sector}) in brackets defines a \emph{flat disk} of radius $\theta_0$, spanned by dimensionless radial coordinate $\theta$ and azimuthal coordinate $\phi$. As can be read directly from (\ref{eq:growing_disk_sector}), the dimensionless disk grows with coordinate $r$. This "growth" of the disk follows directly from conical geometry (see \textbf{Appendix A}).

\begin{figure}[H]
\centering
\includegraphics[width=10.5 cm]{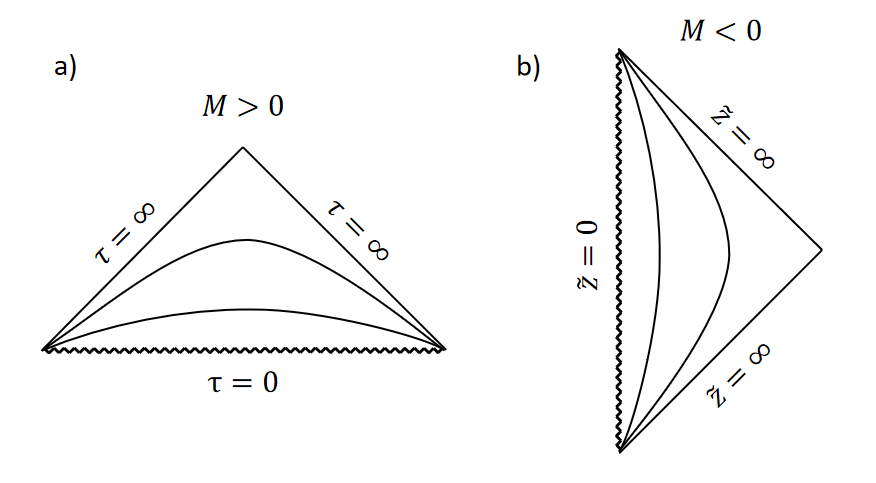}
\caption{Conformal (Penrose) diagrams for the $AIII$ metrics: a) Kasner solution; b) Taub solution. \label{Penrose_pic}}
\end{figure}

It is instructive to compare the Penrose diagram of the type D Kasner solution with that of the inside of a Schwarzschild black hole, as they show the same near-singularity structure. The condition $r \ll 2M$ for the Schwarzschild spacetime is equivalent to $\tau \ll 4M/3$ in Kasner coordinate, therefore, the small-$\tau$ sector of the Kasner Penrose diagram (see left side of Figure \ref{Penrose_pic}) matches the $r \ll 2M$ sector of the Schwarzschild Penrose diagram; in a similar way, the conformal diagram of the Taub solution (right side of Figure \ref{Penrose_pic}) for small-$\tilde{z}$ matches that of negative-mass Schwarzschild for $r \ll 2|M|$ \cite{Griffiths}.

This provides an example of an interesting feature of general relativity, namely that a single metric can describe apparently unrelated physical phenomena: in this case, both the birth and expansion of a universe and the spacetime deep within a black hole. To reiterate, the cosmological picture and the deep-radial Schwarzschild interpretation of the ($M>0$) $AIII$-metric (\ref{eq:AIII_metrics}) are both valid and not mutually exclusive. 

Let us now discuss the case of negative parameter, $M<0$. A simple coordinate transformation (see \textbf{Appendix C}) shows that metric (\ref{eq:metric_neg_M}) is in fact the Taub solution, the only static vacuum solution with planar symmetry \cite{Griffiths,Taub2,Taub1}; this solution has a timelike singularity at $r=0$.

There have been numerous attempts to physically interpret the $AIII$-metric for $M<0$ [Eq. (\ref{eq:metric_neg_M})], but, as stated in Ref. \cite{Griffiths}, "a totally satisfactory interpretation of this simple static metric has still not been found". One of the reasons for this past difficulty may have been a too-strong focus on obtaining a \emph{global} interpretation of the solution, which is problematic due to its planar symmetry. Our approach is local in nature.

Previously-published suggestions include one arguing that the origin at $r=0$ describes a static infinite plane source that repels timelike geodesics \cite{Bedran,Bonnor2,Horsky2}. Another is that this spacetime is a description of the external region sourced by an infinite line of fixed negative gravitational mass per unit length \cite{Griffiths}; yet another interpretation is that the spacetime describes the exterior region of a semi-infinite rod source, again with a specific, fixed negative mass density \cite{Bonnor4}. A reading of the metric as modelling the field due to a null particle \cite{Trautman1} has also been published but has been argued to be incorrect \cite{Bonnor_null,Bonnor4}.

Our new interpretation removes all need for unusual sources (except for negative mass) and fine-tuned parameters, and in addition it explains the behaviour of test particles (timelike geodesics) and null rays as they approach the singularity. The behaviour of geodesics in the Taub spacetime has been studied by Bedran et al. \cite{Bedran}. They found that massive particles \emph{cannot reach the singularity}, independent of initial conditions, but that massless particles can attain the singularity when falling perpendicular to the plane of symmetry. The behaviour of geodesics in negative-mass Schwarzschild spacetime has also previously been studied \cite{Bonnor3,Miller}, demonstrating the same qualitative behaviour: test particles are repulsed from the singularity but a radially-infalling ray of light can reach it in finite time. The consistent behaviour of geodesics in both Taub and negative-mass Schwarzschild spacetimes is not a coincidence, as the former describes a local region of the latter.

Our interpretation also explains a hitherto mysterious feature of the behaviour of test particles in Taub spacetime, pointed out by Bonnor \cite{Bonnor2}: the proper distance between two test particles decreases as the singularity is approached. There is no apparent reason for this if the singular source were a plane, however, our new interpretation resolves this: as the $r=0$ singularity is approached, two neighbouring paths will indeed begin to coalesce for negative-mass Schwarzschild spacetime, as the "Taub plane" itself (the base of the cone in \textbf{Appendix A}) shrinks.

Another piece of evidence against a negative-mass planar source seeding Taub spacetime, noted previously \cite{Griffiths,Bonnor2}, is that general relativity does not seem to predict a corresponding positive-mass planar source, making the negative-mass source suspect. Our interpretation resolves this apparent discrepancy, as both positive- and negative-mass sources are provided by the $M>0$ and $M<0$ Schwarzschild solutions, respectively. This pleasing symmetry in our new reading of the $AIII$-metrics illuminates the previously-noted "dual" nature \cite{Bedran} of the Taub and Kasner solutions.

As has been pointed out by other authors, the Kasner and Taub solutions also emerge from \emph{infinite-mass limits} of Schwarzschild spacetime \cite{Bedran,Ehlers,Trautman1,Geroch,Belinski,Horsky}. A simple proof of this fact was given in Ref. \cite{Horsky}, which we reproduce here. The following solution of Einstein's field equations,
\begin{equation} \label{eq:Horsky_12}
ds^2 = -\left( K - \frac{2M}{r}\right) dt^2 + \left( K - \frac{2M}{r}\right)^{-1} dr^2 + r^2 \left( d\widetilde{\theta}^2 + \mathrm{cos}^2 \left( \sqrt{K} \widetilde{\theta} \right) d\phi^2 \right) ,
\end{equation}
has a limit\footnote{Note the unusual form that the Schwarzschild metric takes here in the $K\rightarrow 1$ limit, due to the presence of the $\mathrm{cos}^2 \widetilde{\theta}$ term, as opposed to the usual $\mathrm{sin}^2 \theta$ term. This is nothing more than a rotation of the coordinate system by 90 degrees: $\widetilde{\theta} \equiv \theta - \pi/2$.} to Schwarzschild spacetime when $K\rightarrow 1$ and a limit to Taub (Kasner) spacetime when $K\rightarrow 0$ and $M<0$ ($M>0$). The symbol $K$ now represents a parameter taking an arbitrary value and no longer denotes the Kretschmann invariant as it did in Sec. \ref{sec:results}. Using the coordinate transformations
\begin{equation}
R\equiv r/ \sqrt{K}, \qquad T\equiv t\sqrt{K}, \qquad \overline{\theta} \equiv \widetilde{\theta} \sqrt{K}, \qquad \overline{\phi} \equiv \phi \sqrt{K},
\end{equation}
metric (\ref{eq:Horsky_12}) becomes
\begin{equation}
ds^2 = -\left( 1 - \frac{2\widetilde{M}}{R} \right) dT^2 + \left( 1 - \frac{2\widetilde{M}}{R} \right)^{-1} dR^2 + R^2 \left( d\overline{\theta}^2 + \mathrm{cos}^2 \overline{\theta} d\overline{\phi}^2 \right) ,
\end{equation}
which is the Schwarzschild solution with mass parameter $\widetilde{M} = M/(K^{3/2})$. As is clear in these new coordinates, for positive $M$, the limit $K\rightarrow 0$ corresponds to the limit $\widetilde{M}\rightarrow \infty$. For negative $M$, the limit $K\rightarrow 0$ corresponds to $\widetilde{M}\rightarrow -\infty$.

We argue that the deep-radial interpretations of the Kasner and Taub metrics are considerably more natural than the ones suggested by the above limits, as the former avoid the pathologies of infinite positive or negative mass, which are evidently unphysical.

In the future, it is hoped that other simple physical pictures for solutions to Einstein's field equations may be found, using local techniques such as ours when global descriptions are lacking. It would be interesting to investigate whether local approaches are also useful in finding new interpretations of other $A$-metrics and of generalisations thereof with multiple parameters.

\section{Conclusions} \label{sec:conclusions}

We have put forward a new interpretation of a member of the $AIII$ class of metrics, namely the plane symmetric static Taub solution. A new, more transparent derivation of the relationship between the region deep within a Schwarzschild black hole and the type D Kasner solution is also presented, highlighting the "duality" between Kasner and Taub solutions. These solutions to Einstein's field equations in vacuum have been known for decades, however, for the Taub case, a natural, agreed-upon physical description has been lacking. We show that, depending on the sign of parameter $M$ (the only parameter in the solution), the $AIII$-metrics describe either a local region deep inside a Schwarzschild black hole or a sector of negative-mass Schwarzschild spacetime. These simple, dual descriptions are compared critically with previously published ones.

\vspace{6mm}

\noindent \textbf{Acknowledgments:} The authors wish to thank Kexin Wang for his help with the figures.
\newpage

\section*{Appendix A: Illustration of the geometries}

If one fixes the variables $t$ and $r$ in metric (\ref{eq:metric_pos_M}) to constants, $t=t_0$ and $r=r_0$, the line element reduces to
\begin{equation} \label{eq:disk_reduced}
ds^2 = {r_0}^2 \left( d\theta^2 + \theta^2 d\phi^2 \right) .
\end{equation}
We now show that this is a simple example of conical geometry, with the first fundamental form (\ref{eq:disk_reduced}) describing the base of a cone for small apex angle, thereby proving our assertion in the main text that $\theta$ can be treated as a dimensionless radial coordinate after enforcing the constraint $\theta \ll 1$, and making the geometry easier to visualise.

\begin{figure}[H]
\centering
\includegraphics[width=6.5 cm]{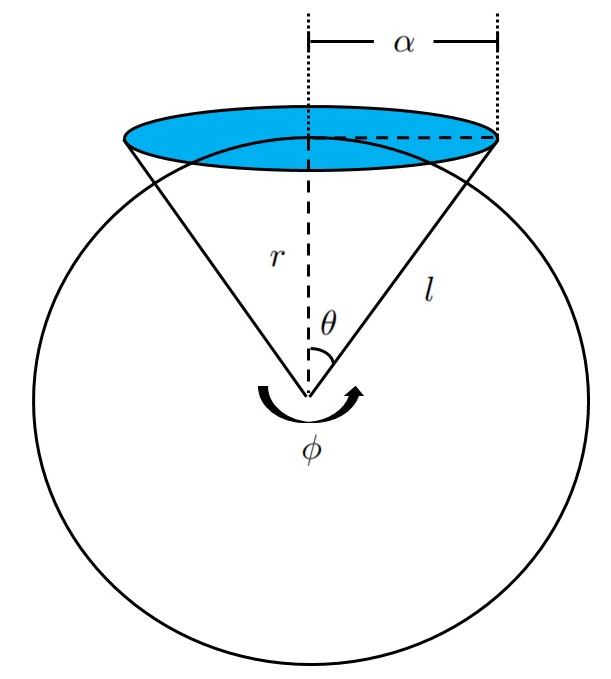}
\caption{A schematic of a cone embedded into a sphere, illustrating the approximation of a small region of a two-sphere as a flat disk. The relative size of the disk (blue) has been exaggerated for clarity. \label{cone_pic}}
\end{figure}

A right circular cone can fit into a spherical coordinate system as shown in Figure~\ref{cone_pic}; the Schwarzschild coordinates $\theta$ and $r$ coincide with the apex angle and the height of the cone, respectively. The azimuthal angle $\phi$ is both the standard Schwarzschild azimuth and that defining the polar angle of the disk/base of the cone. We denote the slant height and radius of the cone by $l$ and $\alpha$, respectively.

Constraining the apex angle of the cone to be small, $\theta \ll 1$, the following relations hold: $\mathrm{tan} \theta = \alpha / r$ (in general) and $\mathrm{tan} \theta \approx \theta$, together giving $\alpha \approx r\theta$. Fixing $r=r_0$, one also has $\alpha \approx r_{0} \theta$ and $d\alpha \approx r_{0} d\theta$. The line element describing the base of the cone is then $ds^2 = d\alpha^2 + \alpha^2 d\phi^2 \approx r_{0}^2 \left ( d\theta^2 + \theta^2 d\phi^2 \right)$, which is precisely (\ref{eq:disk_reduced}).

For parameter range $M>0$, this geometry is physically interpreted in the cosmological picture \cite{Griffiths} as a spacetime originating in a big bang (the apex of the cylinder), expanding uniformly in the $x$ and $y$ directions (covering the cylinder's base), and contracting in $z$ (this "fourth" dimension cannot be shown in our schematic); remember that $r$ is timelike in this case, where Figure~\ref{cone_pic} represents a cosmology. For $M<0$, $r$ is spacelike.

\section*{Appendix B: Coordinate transformation to the type D Kasner solution and spaghettification}

The coordinate transformation putting Eq. (\ref{eq:AIII_metrics}) in the form of Eq. (\ref{eq:Kasner}), i.e. demonstrating that the $M>0$ $AIII$-metric is in fact the type D Kasner vacuum solution, is as follows \cite{Griffiths}:
\begin{equation}
t \equiv \left( \frac{3}{4M} \right)^{1/3} z, \qquad r \equiv \left( \frac{9M}{2} \right)^{1/3} \tau^{2/3}, \qquad
\rho e^{i\phi} \equiv \left( \frac{2}{9M} \right)^{1/3} \left( x+iy \right) .
\end{equation}
Note that the labelling after the transformation reflects more plainly the nature (timelike or spacelike) of the coordinates.

The specific form of the above coordinate transformation sheds much light on the interpretation of the Kasner metric as a region of the Schwarzschild solution. The radial transformation\footnote{This relation between $r$ and $\tau$ is also used to define the Lema\^{i}tre reference frame \cite{Frolov}.}, i.e. $r \equiv \left( 9M / 2 \right)^{1/3} \tau^{2/3}$, is also precisely the relation between the radial coordinate of an observer falling into a Schwarzschild black hole and the proper \emph{countdown time} $\tau$ until they reach the singularity $r=0$ \cite{MTW,Lambourne}. (The observer is assumed to fall inwards with purely radial motion and to begin the fall at a large distance from the black hole.) The countdown time $\tau$ is defined as the proper time at which the singularity will be reached minus the proper time measured by the infalling observer. As the time $\tau$ ticks down in metric (\ref{eq:Kasner}) (i.e. as $r=0$ is approached), the spacetime can be seen to extend along $z$ and contract circumferentially in $x$ and $y$ at the rate known for an infalling "spaghettified" observer (see p. 862 of Ref. \cite{MTW}).

The cosmological picture discussed in the main text is in a sense the time-reversal of the deep-radial interpretation, as the big bang and growth of a universe in the cosmological picture is, once the "direction" of coordinate $\tau$ is inverted, the spacetime experienced by an observer falling towards the singularity of a black hole.

\section*{Appendix C: Coordinate transformation to the Taub solution}

The coordinates of metric (\ref{eq:metric_neg_M}) can be transformed as \cite{Griffiths}
\begin{equation}
t \equiv |M|^{-1/3} \ \tilde{t}, \qquad r \equiv 2|M|^{1/3} \ \tilde{z}^{1/2}, \qquad
\rho e^{i\phi} \equiv \frac{1}{2}|M|^{-1/3} \left( \tilde{x} + i\tilde{y} \right) ,
\end{equation}
yielding
\begin{equation}
ds^2 = \tilde{z}^{-1/2} \left( -d\tilde{t}^2 + d\tilde{z}^2 \right) + \tilde{z} \left( d\tilde{x}^2 + d\tilde{y}^2 \right) ,
\end{equation}
the Taub solution, the only plane-symmetric static vacuum solution of Einstein's field equations.

\end{document}